\documentclass[epj]{svjour}
\usepackage{graphics}

\begin{document}
\title{Rapidity bin multiplicity correlations from a multi-phase transport model}
\author{Mei-Juan Wang\inst{1}
\and 
Gang Chen\inst{1}
\and
Yuan-Fang Wu\inst{2}
\and
Guo-Liang Ma\inst{3}}

\mail{wangmj@cug.edu.cn, glma@sinap.ac.cn}

\institute{School of Mathematics and Physics, China University of Geoscience, Wuhan 
430074, China
\and
Key Laboratory of Quark and Lepton Physics (MOE) and Institute of Particle 
Physics, Central China Normal University, Wuhan 430079, China
\and
Shanghai Institute of Applied Physics, Chinese Academy of Sciences, Shanghai 201800, China}

\date{Received: date / Revised version: date}

\abstract{The central-arbitrary bin and forward-backward bin multiplicity correlation patterns 
for Au+Au collisions at $\sqrt{s_{NN}} $= $7.7-62.4$ GeV 
are investigated within a multi-phase transport (AMPT) model.
An interesting observation is that for $\sqrt{s_{NN}} <19.6$ GeV Au+Au collisions,  
these two correlation patterns both have an increase with the pseudorapidity gap,  
while for $\sqrt{s_{NN}} >19.6$ GeV Au+Au collisions, they decrease.
We mainly discuss the influence of different evolution stages of collision system on the central-arbitrary 
bin correlations, such as the initial conditions, partonic scatterings, 
hadronization scheme and hadronic scatterings. Our results show that 
the central-arbitrary bin multiplicity correlations have different responses to partonic phase and hadronic phase, 
which can be suggested as a good probe to explore the dynamical evolution mechanism 
of the hot dense matter in high-energy heavy-ion collisions.
}

\authorrunning
\titlerunning
\maketitle

\section{Introduction}
The main goal of high-energy heavy-ion collisions is to create a hot dense matter, called 
Quark-Gluon Plasma (QGP)~\cite{QGP1,QGP2},  and to study its properties.  There are convincible evidences 
to prove that the QGP or quark matter has been produced at RHIC energy,  such as the quark scaling of 
elliptic flow~\cite{flow-scaling-ex,flow-scaling-th}, the energy loss of jet quenching~\cite{energy-loss-ex,energy-loss-th}.  
The properties of QGP have been widely studied by comparing the observables of final state particles 
between A+A collisions and h+h collisions.  
In this way, main differences of particle production are expected for the two conditions 
with or without the QGP~\cite{com-pp-AuAu1,com-pp-AuAu2}.  

Among the different observables, such as the anisotropic flow~\cite{flow-first1,flow-first2} and 
nuclear modification factor $R_{AA}$~\cite{Raa-first},  
the multiplicity correlations between different regions of rapidity are considered to be a good probe 
to investigate the new state of matter~\cite{correlation1,correlation2}. 
In h+h collisions, the correlations among particles produced in different pseudorapidity regions had been 
studied for the dynamics of particle production in 1980s-1990s~\cite{1986prd,1994prl}.  
In A+A collisions, the longitudinal correlations of final particles are sensitive to 
the evolution mechanism of the collision system, 
and especially have the potential to probe the early states of heavy ion collisions~\cite{FB-ex,FB-th}. 
If a deconfined phase of quarks and gluons exists in these high-energy heavy-ion collisions,  
the presence of the partonic degree freedom could have a direct influence on the correlation measurement, 
{\emph {e.g.}}, a narrowed balance function would imply a delayed hadronization due to a long-lived QGP~\cite{narrow-bf1,narrow-bf2}. 
Therefore, the rapidity correlations can serve as an effective way to explore the properties of 
quark matter produced in high-energy heavy-ion collisions. 

The decorrelation of anisotropic flows with large pseudorapidity gap is 
recently found to be sensitive to the initial condition and dynamical evolution 
of the QGP~\cite{decorrelation1,decorrelation2}. 
One popular method to study the rapidity correlations is to measure the multiplicity 
correlation coefficient, {\emph {i.e.}}, quantify how multiplicity (number of particles) in one rapidity window influences 
multiplicity in another one~\cite{two-rapidity-window}.
It is expected that the QGP is easily formed in the mid-rapidity region in the central A+A collisions
where the energy density is extremely high. 
In this study, we focus on the correlations of particles between central rapidity bin and other rapidity bin, 
called central-arbitrary bin correlation pattern in the following, 
which are considered to "remember" and carry some important information on the QGP properties.  
By way of comparison, we also measure another two-bin correlation pattern, for a symmetric choice of rapidity bin, 
called forward-backward bin correlation pattern. 

In relativistic heavy ion collision experiments, we can only record the physical information of final state particles, 
but can not trace the intermediate evolution process of the collision system. 
Some theoretical tools ~\cite{hydro1,hydro2,xu1,xu2,yan} are essential to better understand 
the dynamical evolution mechanism of collision system. 
A multi-phase transport model (AMPT), in which both partonic and hadronic phases are included, 
is suggested as a great tool for studying the space-time evolution of collision system~\cite{ampt1,ampt2}, 
especially for the case with the QGP production. 
Here, we utilize the AMPT model to simulate a complete evolution process 
from partonic to hadronic phase for Au+Au collisions at $\sqrt{s_{NN}} $= $7.7-62.4$GeV. 
The purpose of our work is to study the central-arbitrary bin multiplicity correlation pattern
for different evolution stages of collision system within the AMPT model in scenario 
(a) the effects of initial state; (b)the early partonic phase;
(c)the intermediate hadronic phase, the time when the collision system 
undergoes the hadronization, while hadronic scattering scheme is turned off; 
(d)the final hadronic phase, when hadronization has happened and 
hadronic scattering scheme is turned on. 
These can present a clear image of correlations for different stages of the system evolution, 
especially for partonic phase and hadronic phase.

This paper is organized as follows.  A short introduction to the AMPT model is given in Section 2.  
In Section 3, we present the AMPT results about the two correlation patterns of final particles 
for Au+Au collisions at $\sqrt{s_{NN}} $= $7.7-62.4$GeV. Then, the central-arbitrary bin 
correlation pattern is studied for three centrality classes,
accounting for 0-10$\%$, 30-40$\%$ and 50-80$\%$. 
In addition, we discuss the effects of both the partonic and hadronic evolutions. 
But we only take $\sqrt{s_{NN}} $= 7.7GeV and 62.4GeV as two examples to 
focus on the influences of partonic evolution on the correlation pattern. 
Finally, some conclusions are given in Section 4. 

\section{A brief introduction to the AMPT model}
The Monte Carlo event generator AMPT (A Multi-Phase Transport) has been used in this study.  
The AMPT model is made up of four main components:  initial conditions,
partonic interactions, hadronization and hadronic interactions. The initial conditions, which include 
the spatial and momentum distributions of minijet partons from hard processes and strings from soft 
processes, are obtained from the Heavy Ion Jet Interaction Generator (HIJING) model~\cite{HIJING1,HIJING2,HIJING3,HIJING4}. 
The evolution of parton phase is modeled by Zhang's Parton Cascade (ZPC)~\cite{ZPC},  which includes only 
parton-parton elastic scatterings with cross sections obtained from the pQCD calculation with screening masses.  
The AMPT model has two versions, the default AMPT model and the AMPT model with string melting mechanism.
In the default AMPT model, only minijet  partons from the initial conditions take part in the interactions modeled by ZPC. 
When stoping interactions, they are combined with their parent strings to form new strings.  
The resulting strings are then converted to hadrons according to a Lund string fragmentation 
model~\cite{fragmentation1,fragmentation2}. 
In the AMPT model with string melting mechanism, the parent strings first fragment into partons and then 
enter the ZPC model together with the minijet partons. 
After freezing out,  a simple quark coalescence model is used to combine the two nearest partons into a meson 
and three nearest partons into a baryon~\cite{coalescence}.  
For both versions of the AMPT model, the interactions among resulting hadrons are described 
by a relativistic transport (ART) model~\cite{ART1,ART2}.  

Compared with the default AMPT model, the partonic phase can be better modeled by the AMPT model with 
string melting mechanism. Because the QGP is expected to be formed in heavy-ion collisions at the Relativistic Heavy-Ion 
Collider(RHIC),  the AMPT with string melting mechanism is considered a more efficient tool to 
study the properties of the new matter, {\emph{e.g.}}, studying the elliptic flow and triangular flow~\cite{flow1,flow2}. 
Based on this, we utilize the AMPT model with string melting mechanism to generate Au+Au collisions 
at $\sqrt{s_{NN}} = 7.7-62.4$ GeV.  The parton cross section is taken to be 10 mb in our simulations.

\begin{figure}
\begin{center}
\resizebox{0.6\hsize}{!}{\includegraphics*{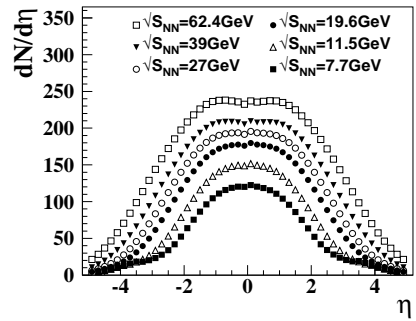}}
\end{center}
\caption{The pesudorapidity distribution of final particles at $\sqrt{s_{NN}}$ =$7.7-62.4$GeV Au+Au collisions 
by using the AMPT model with string melting mechanism. }
\label{Fig1}
\end{figure}

\begin{figure*}
\begin{center}
\resizebox{0.7\hsize}{!}{\includegraphics*{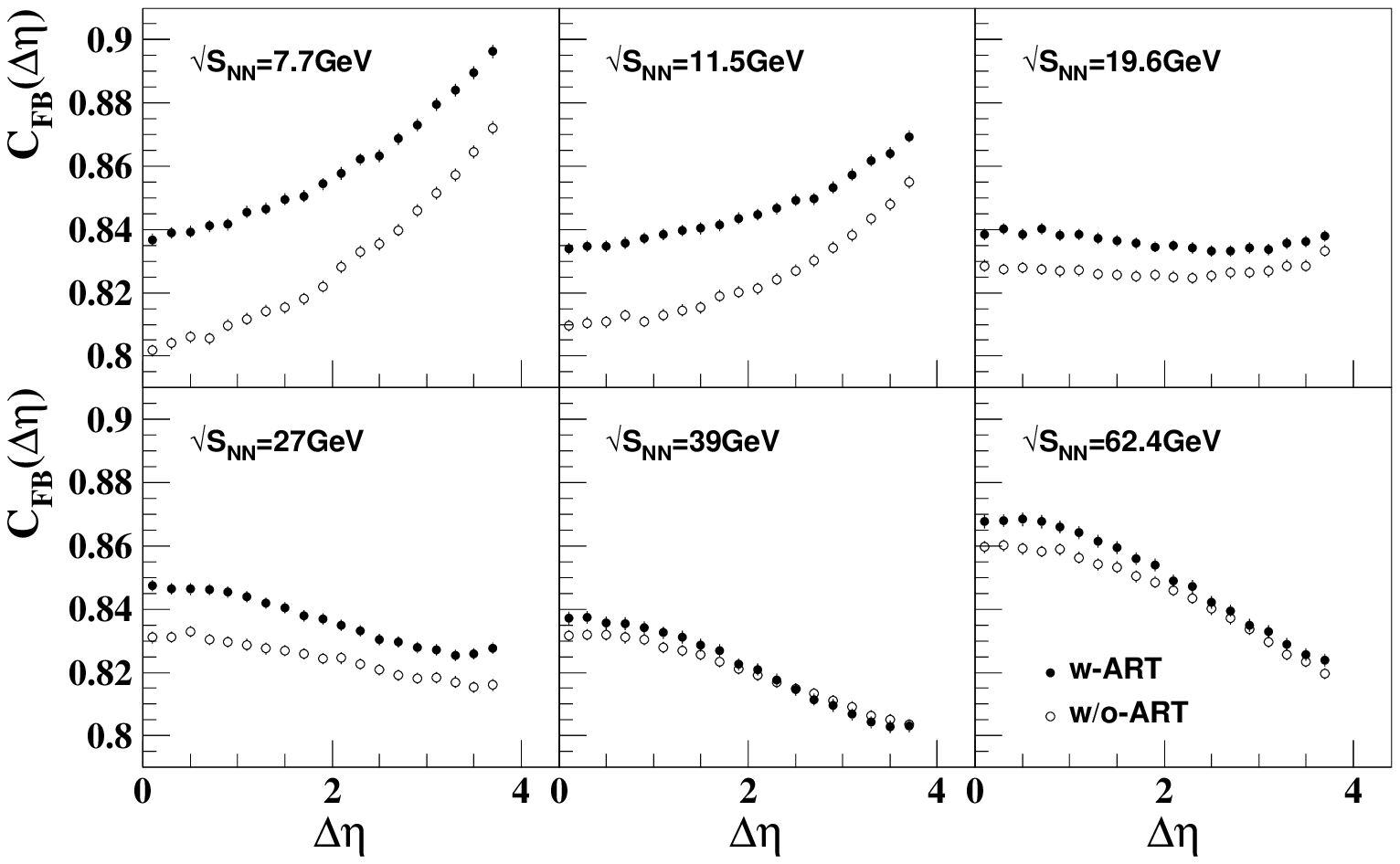}}
\end{center}
\caption{ The pseudorapidity gap dependence of the forward-backward bin multiplicity correlation patterns 
for Au+Au collisions at $\sqrt{s_{NN}}$ =$7.7-62.4$GeV  by using the AMPT model with string melting mechanism, 
where the solid points and open points correspond to the two cases with and without hadronic scatterings, respectively. }
\label{Fig2}
\end{figure*}

\begin{figure*}
\begin{center}
\resizebox{0.7\hsize}{!}{\includegraphics*{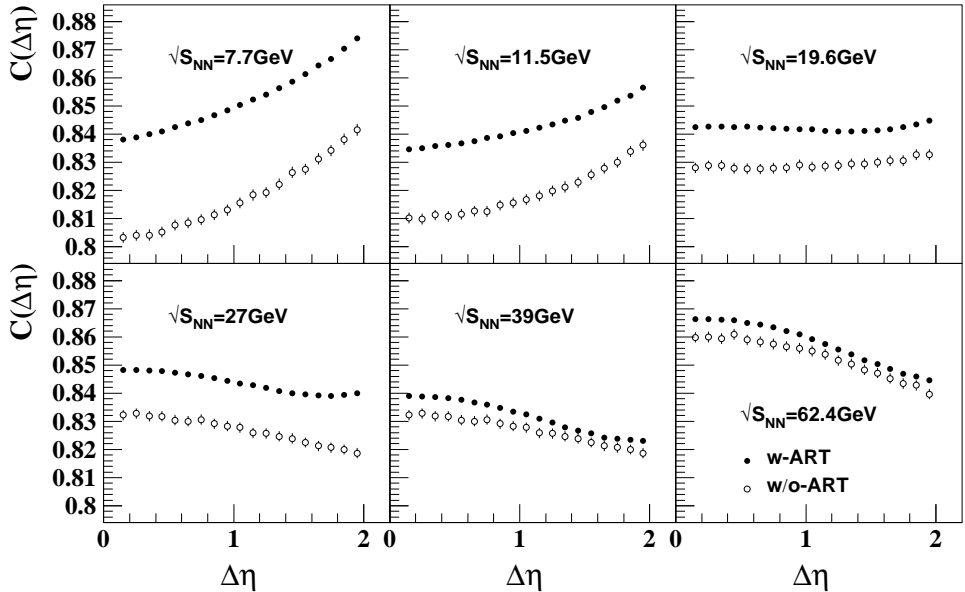}}
\end{center}
\caption{ The pseudorapidity gap dependence of the central-arbitrary bin multiplicity correlation patterns 
for Au+Au collisions at $\sqrt{s_{NN}}$ =$7.7-62.4$GeV by using the AMPT model with string melting mechanism, 
where the solid points and open points correspond to the two cases with and without hadronic scatterings, respectively. }
\label{Fig3}
\end{figure*}

\section{Pseudorapidity bin multiplicity correlations }
One focus of the analysis of the final state is on the longitudinal momentum distributions, and correlations. 
In this section, we first show the variation in particle density with $\eta$. 
Then, we mainly measure the correlation patterns as a function of pseudorapidity gap at different colliding energies, 
and especially discuss the influences of partonic and hadronic scatterings on the correlation patterns.

In Fig.~\ref{Fig1}, we present  the pseudorapidity distributions of final particles within -5 $\le$ $\eta$ $\le$ 5 
for Au+Au collisions at $\sqrt{s_{NN}} $= $7-62.4$GeV.  
As expected, the particle density increases with decreasing $|\eta|$ for all energies. 
As a function of collision energy, the pseudorapidity distribution grows systematically both in height and width. 
In the following, we study the pseudorapidity correlations of final particles to obtain a deeper 
insight into particle production mechanisms. 

\begin{figure*}
\begin{center}
\resizebox{0.8\hsize}{!}{\includegraphics*{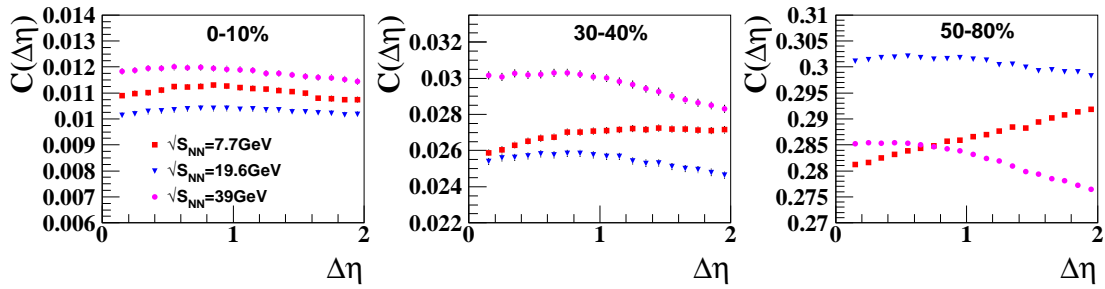}}
\end{center}
\caption{ The pseudorapidity gap dependence of the central-arbitrary bin multiplicity correlation patterns 
for 0-10$\%$, 30-40$\%$ and 50-80$\%$ Au+Au collisions at $\sqrt{s_{NN}} $= 7.7GeV, 19.6GeV and 39GeV by using the AMPT model 
with string melting mechanism.}
\label{Fig4}
\end{figure*}

The general two-bin multiplicity correlation is defined as 
\begin{eqnarray}
C_{1,2}=\frac{\langle
n_{1}n_{2}\rangle}{\langle n_{1}\rangle\langle n_{2}\rangle}-1.
\end{eqnarray}
where $n_{1}$ and  $n_{2}$ denote the multiplicities in bin $1$ and bin $2$, respectively. 
If particles are produced independently over the whole phase space, 
then$ <n_{1}n_{2}>=<n_{1}><n_{2}> $ and $C_{1,2}$ vanishes.   

In this way, we can obtain a three-dimensional correlation pattern by varying bin $1$ and bin $2$. 
However, the structure of this kind of three-dimensional correlation pattern is complicated 
and it is not so easy to gain the information on the underlying dynamics intuitively. 
A construction of two-dimensional pattern is necessary to show the fine structure more clearly. 
Now we introduce two methods to construct a two-dimensional correlation pattern. 

One is choosing two bins symmetrically with $\eta =0$, called forward-backward bin multiplicity correlation pattern, 
which can give information about the earliest stage of high-energy heavy-ion collisions~\cite{long-range-early}. 
Based on Eq. 1, the forward-backward bin multiplicity correlation pattern can be 
proposed as 

\begin{eqnarray}
C_{FB}(\Delta \eta)=\frac{\langle
n_{\eta_F}n_{\eta_B}\rangle}{\langle n_{\eta_F}\rangle\langle n_{\eta_B}\rangle}-1.
\end{eqnarray}
where $\eta_{F}$ and $\eta_{B}$ are located symmetrically about midrapidity ($\eta =0$) 
with a pseudorapidity gap $\Delta \eta$. 
$n_{\eta_{F}}$ and $n_{\eta_{B}}$ correspond to the multiplicities in the forward
hemisphere and backward hemisphere, respectively.  
We divide the pseudorapidity region  [-2, 2] equally 
into 20 bins,  which corresponds to the bin width  $\delta \eta =0.2$. 

The other is fixing one bin and varying the other bin, called fixed-to-arbitrary bin multiplicity correlation pattern, 
which is proved to be efficient to identify various random multiplicative cascade processes~\cite{wu-pre}. In this 
work, we apply it to high-energy heavy-ion collisions. The QGP is thought to be produced in the mid-rapidity region 
where the energy density is extremely high.  Based on this point, we choose the central rapidity as the fixed bin, 
and study the correlations between the particles at the central rapidity bin and those in other mid-rapidity bin
to track the information on the QGP properties. 
Motivated by these assumptions, the central-arbitrary bin multiplicity correlation pattern can be defined as 

\begin{eqnarray}
C(\Delta \eta)=\frac{\langle
n_{\eta_0}n_{\eta}\rangle}{\langle n_{\eta_0}\rangle\langle n_{\eta}\rangle}-1.
\end{eqnarray}

In this definition, $\eta_{0}$ corresponds to the central rapidity bin, and $\eta$ varies 
from -2 to 2 for the different pseudorapidity gap $\Delta \eta$. $n_{\eta_{0}}$ and $n_{\eta}$ 
are the multiplicities in the central rapidity bin and other rapidity bin, respectively.
Note we only present the correlation pattern in the positive $\Delta \eta$ direction in this paper, 
because of the symmetry that $C(\Delta \eta)=C(-\Delta \eta)$ in A+A collisions. 

\begin{figure*}
\begin{center}
\resizebox{0.7\hsize}{!}{\includegraphics*{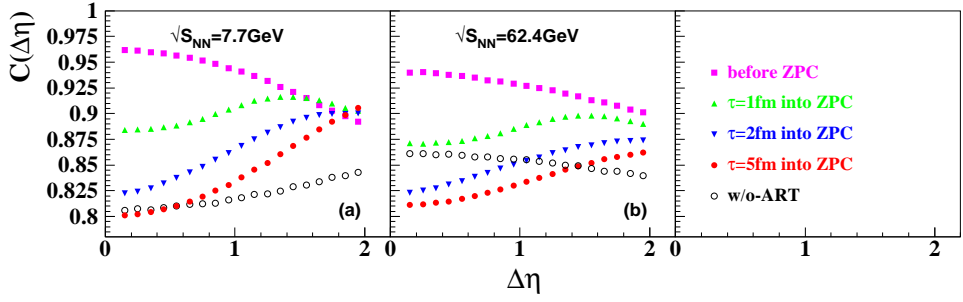}}
\end{center}
\caption{ The pseudorapidity gap dependence of the central-arbitrary bin multiplicity correlation pattern 
for different evolution stages of Au+Au collisions at $\sqrt{s_{NN}} $= 7.7GeV and 62.4GeV by using the AMPT model 
with string melting mechanism. }
\label{Fig5}
\end{figure*}

The forward-backward bin and central-arbitrary bin correlation patterns for Au+Au collisions 
at $\sqrt{s_{NN}} = 7.7-62.4$ GeV are shown in Fig. 2 and Fig. 3, respectively.
We calculate two cases of correlations, 
{\emph{i.e. }}, (a) ``w-ART", which represents the results from a complete time evolution process of the AMPT model, 
 shown as solid circles; 
 (b) ``w/o- ART" represents the results from the time just after quark coalescence but before hadronic scatterings,  
 shown as open circles.  
 By comparing these two cases, it can help us to understand the hadronic effect on the correlation patterns. 
 Both for the two figures, the results from ``w/o-ART" are lower than those from ``w-ART" for all collision energies, 
 which means hadronic scatterings can increase the correlation strength to some level.  
 In addition, the values of  ``w/o-ART" are closer to those of ``w-ART" at higher energies.  
 This indicates that the influence of hadronic interactions on correlation patterns becomes 
 weaker with increasing energy.  

By comparing the Fig.~\ref{Fig2} and Fig.~\ref{Fig3},  another interesting phenomenon is 
that the two correlation patterns show a similar tendency varying with the pseudorapidity gap 
for the same energy range. However, for different energy ranges,  both the two correlation 
patterns have quite different dependences with pseudorapidity gap. In particular, 
for $\sqrt{s_{NN}} =7.7$  and 11.5 GeV Au+Au collisions, the correlation values increase 
with increasing pseudorapidity gap, while for $\sqrt{s_{NN}} =27$, 
39 and 62.4 GeV Au+Au collisions, they decrease.  For $\sqrt{s_{NN}}$ = 19.6GeV Au+Au collisions, 
the correlation values almost remain unchanged with increasing pseudorapidity gap. 
Considering the similar pseudorapidity gap dependence of the two correlation patterns, 
we only focus on the central-arbitrary bin correlation pattern. 

In order to eliminate (or at least reduce) the effect of the centrality fluctuations, 
we measure the correlation pattern in a narrow centrality interval.  As an example, 
Au+Au collisions at $\sqrt {s_{NN}} $= 7.7GeV, 19.6GeV and 39GeV are chosen to 
study the centrality effect. The centralities studied in 
this analysis account for 0-10$\%$, 30-40$\%$ and 50-80$\%$. 
The central-arbitrary bin correlation patterns for the three centrality classes are shown in Fig.~\ref{Fig4}. 
From this figure, we can see a major increase for the correlation values from central to peripheral collisions. This indicates 
the large magnitude of correlation pattern in Fig.~\ref{Fig3} mainly comes from 
event-by-event fluctuations. In addition, for mid-central and peripheral Au+Au collisions, we again obtain that the correlation pattern 
has an increase for $\sqrt {s_{NN}} $= 7.7GeV and an decrease for $\sqrt {s_{NN}} $= 39GeV with increasing pseudorapidity gap. 
A possible explanation of correlation behavior for peripheral $\sqrt {s_{NN}} $= 7.7GeV Au+Au collisions is the 
long-range effect caused by energy conservation because of low multiplicity. 

To further explore the origin of all the observed correlation behaviors for different energies 
and different centralities, let's go back to the Fig.~\ref{Fig3}. 
Another concern is that hadronic interactions can increase the 
correlation values at low energies, and the influence becomes negligible at high energies. 
As we know, the AMPT model is a hybrid model, in which both partonic and hadronic interactions 
are included~\cite{ampt1,ampt2}. Because of the QCD phase diagram,  hadronic degrees of freedom turn out 
to be important at low energies, while partonic degrees of freedom play a key role at high energies~\cite{Nu-QCD}. 
It is widely believed that the hadronic interactions dominate the correlation behaviors at low energies 
and partonic interactions dominate the correlation behaviors at high energies. 
Based on this, further study of partonic evolution on the correlation measurement is 
needed to fully understand the observed effect.

Next, Au+Au collisions at $\sqrt {s_{NN}} $= 7.7GeV and 62.4GeV are chosen 
to study the correlation patterns for partonic phase, since they 
correspond to two cases for low energy and high energy, respectively.  
In Fig.~\ref{Fig5},  we present the correlation patterns of partons with time evolution for the two energies. 
Four important evolution times are considered.  
The ``before ZPC" represents the parton correlation pattern from the initial state of partonic matter;  
the "$\tau=1 fm/c$",  "$\tau=2 fm/c$" and "$\tau=5 fm/c$" denote the correlation patterns 
at three time points when the partonic evolution has been going on 
for 1 $fm/c$, 2 $fm/c$ and 5 $fm/c$, respectively. 
To see the hadronization effect, the result of "w/o-ART" is 
plotted to enable a visual comparison.

In Fig.~\ref{Fig5}, we can see that the correlation patterns decrease with the pseudorapidity gap
 both for the initial state of Au+Au collisions at $\sqrt {s_{NN}}$ = 7.7GeV and 62.4GeV  ({\emph{i.e.}}, ``before ZPC").  
 It is consistent with the HIJING results that show strong short-range correlations~\cite{HIJING-correlation}.  
 When partons take part in the process of parton cascade,  the correlation patterns become an increasing trend 
 with the pseudorapidity gap, because the short-range correlations are strongly weaken 
 but the long-range correlations tend to persist (or weakly weaken) with the evolution time of parton cascade. 
 The weaken effect of correlation pattern mainly comes from the strong partonic scatterings.  
 In addition, we note that the correlation patterns of  ``$\tau=1 fm/c$", ``$\tau=2 fm/c$" and ``$\tau=5 fm/c$" 
 intersect at near $|\Delta \eta|=2$ for $\sqrt {s_{NN}}$ = 7.7 GeV Au+Au collisions, 
 while the intersection point is expected at a larger pseudorapidity gap for $\sqrt {s_{NN}}$ = 62.4 GeV Au+Au colllisions.  
 We argue that it is probably because the QGP production in the mid-rapidity region can spread the correlations of partons 
 to a larger space for $\sqrt{s_{NN}}$ = 62.4GeV Au+Au collisions.  

\begin{figure}
\begin{center}
\resizebox{0.6\hsize}{!}{\includegraphics*{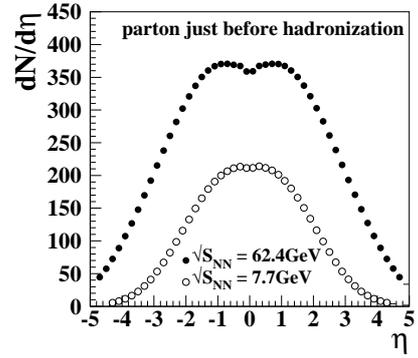}}
\end{center}
\caption{The pseudorapidity distribution of partons 
right before hadronization for $\sqrt{s_{NN}} $= 7.7GeV (open circle) and 62.4GeV (solid circle) 
Au+Au collisions by using the AMPT model with string melting mechanism.}
\label{Fig6}
\end{figure}

After partons in the string melting scenario cease their interactions, their
hadronization is modeled via a simple quark coalescence~\cite{coalescence}. 
In the quark coalescence model, we combine the two nearest partons into a meson 
and three nearest quarks into a baryon.  
By combining the nearest quarks into hadrons, this hadronization scheme can obviously 
increase short-range correlations and the increasing degrees 
depend on the parton density right before hadronization. 
In Fig.~\ref{Fig6}, we plot the pseudorapidity distribution of partons right before hadronization
both for $\sqrt {s_{NN}}$ = 7.7GeV and 62.4GeV Au+Au collisions. 
The parton density at mid-rapidity for Au+Au collisions at $\sqrt {s_{NN}}$ = 7.7GeV has only 
around half of that at $\sqrt {s_{NN}}$ = 62.4GeV. 
Therefore, for $\sqrt {s_{NN}}$ = 7.7GeV Au+Au collisions, 
a limited number of partons passing through hadronization only have a weak influence on the short-range 
correlations. After taking into account the long-range effect caused by the energy conservation 
at small collision energies, 
the correlation pattern keeps an increase with the pseudorapidity gap after hadronization, 
as shown by open circles in Fig.~\ref{Fig5}(a). While for $\sqrt {s_{NN}}$ = 62.4GeV Au+Au collisions, 
more partons can lead to stronger short-range correlations during the hadronization process. 
It is understandable that the slowly increasing correlation pattern during partonic evolution 
becomes a decreasing trend with the pseudorapidity gap after hadronization, 
as shown by open circles in Fig.~\ref{Fig5}(b).

\section{Conclusions}
In summary, the forward-backward bin and central-arbitrary 
bin multiplicity correlation patterns in the pseudorapidity phase are studied for Au+Au collisions 
at $\sqrt{s_{NN}} $= $7.7-62.4$ GeV within the AMPT model.  
We found that for $\sqrt{s_{NN}} <19.6$ GeV Au+Au collisions, these two correlation patterns 
both have an increase with increasing pseudorapidity gap, 
while for $\sqrt{s_{NN}} >19.6$ GeV Au+Au collisions, they decrease.  
In this paper, our study focuses on the central-arbitrary bin multiplicity 
correlation pattern. 

On one hand, to reduce the impact parameter fluctuations, 
we measure the correlation patterns
at $\sqrt {s_{NN}} $= 7.7GeV, 19.6GeV and 39GeV Au+Au collisions 
in a narrow centrality interval, such as 0-10$\%$, 30-40$\%$ and 50-80$\%$. 
We observe that the correlation values have an overall increase 
from central to peripheral collisions, which can explain the large magnitude 
of correlation values for mini-bias Au+Au collisions as a result of event-by-event 
fluctuations. 

One the other hand, we mainly discuss the influence of dynamical evolution of collision system 
on the correlation patterns for Au+Au collisions at $\sqrt {s_{NN}} $= 7.7GeV and 62.4GeV.
By comparing the AMPT results with and without hadronic scatterings, 
we obtain that hadronic scatterings can increase the correlation values
to some level, and the influence becomes negligible at high energies.  
Further study shows that the correlation patterns increase with 
pseudorapidity gap in the beginning of partonic evolution both for 
$\sqrt {s_{NN}} $= 7.7GeV and 62.4GeV  Au+Au collisions. 
The result demonstrates that the partonic scatterings can weaken the short-range correlations 
from the initial state of collision system and tend to persist the long-range correlations. 
In addition, it is noted that 
the hadronization scheme --- quark coalescence, also has an influence on the correlation pattern. 
The quark coalescence mechanism, which combines the nearest two (three) quarks into a meson (baryon), 
can obviously enhance the short-range correlations. 
By combining the opposite effects of partonic scatterings and the following quark coalescence 
mechanism on correlation patterns, it is understandable that 
the correlation patterns have an increase or a decrease, or even remain nearly 
constant with pseudorapidity gap for different centralities at different energy regimes. 
It is also worth mentioning that a long-range effect caused by energy conservation 
can not be ignored for peripheral Au+Au collisions at low energies, 
where the multiplicities are not very high. 

Since the choice of energy is in connection with the STAR BES program, 
these phenomenal results can extend our understanding about the 
dynamical mechanism of system evolution 
in high-energy heavy-ion collision experiments. We argue that the 
central-arbitrary bin and forward-backward bin multiplicity pseudorapidity correlations, 
which have different responses to partonic phase and hadronic phase, will probably be 
of interest to the future experiments. 

\begin{acknowledgement}
The author Mei-Juan Wang thanks Prof. C. M. Ko, 
Dr.  Ming-Mei Xu, You Zhou and Zi-Qiang Zhang for useful discussions. 
This work was supported in part by GBL31512, the Major State Basic 
Research Development Program in China (No. 2014CB845402), 
the NSFC under Grant Nos.11475149, 11522547, 
11375251, 11421505 and 11221504.
\end{acknowledgement}

\end{document}